\documentclass[a4paper,11pt]{article}
\usepackage[utf8x]{inputenc}
     
\usepackage{a4wide}
\usepackage{amssymb}
\usepackage{amsmath}
\usepackage{amsthm}
\usepackage[mathscr]{euscript}
\usepackage[colorlinks=true]{hyperref}
\usepackage[usenames,dvipsnames]{xcolor}
\usepackage{todonotes}

\numberwithin{equation}{section}

\theoremstyle{definition}
\newtheorem{definition}{Definition}[section]

\theoremstyle{plain}
\newtheorem{Theorem}[definition]{Theorem}

\theoremstyle{remark}

\newcommand{\R}{\mathbb R}

\newcommand{\eps}{\varepsilon}

\usepackage[xcolor]{changebar}
\cbcolor{blue}

\newcommand{\sse}{\subseteq}

\usepackage[inline]{enumitem}
\newcommand{\enumlabelformat}{\roman}
\newcommand{\enumlabelfont}[1]{#1}
\newlength{\thelabelsep}
\setlist{labelsep=\thelabelsep}
\setlist[enumerate]{font=\enumlabelfont,label=(\enumlabelformat*),leftmargin=2.5em}
\setlist[itemize]{leftmargin=2.5em,label=$-$}

\newcounter{inlineenum}
\renewcommand{\theinlineenum}{\enumlabelformat{inlineenum}}

\title{Maximizers in Lipschitz spacetimes are either timelike or null}

\author{Melanie Graf\footnote{University of Vienna, Faculty of Mathematics,
melanie.graf@univie.ac.at}, \\
Eric Ling\footnote{University of Miami, Department of Mathematics, eling@math.miami.edu}
}

\begin{document}

\date{\today}


\maketitle

\begin{abstract} 
We prove that causal maximizers in $C^{0,1}$ spacetimes are either timelike or null. This question was posed in \cite{SS} since bubbling regions in $C^{0,\alpha}$ spacetimes ($\alpha <1$) can produce causal maximizers that contain a segment which is timelike and a segment which is null, cf.~\cite{CG}.
While $C^{0,1}
$ spacetimes do not produce bubbling regions, the
causal character of maximizers for spacetimes with regularity at least  $C^{0,1}$ but less than $C^{1,1}$ was unknown until now. As an application we show that timelike geodesically complete spacetimes are $C^{0,1}$-inextendible.


\vskip 1em

\noindent
\medskip

\noindent

\end{abstract}

\section{Introduction}\label{sec:intro}


Recently, there has been an interest in low regularity aspects of general relativity
 motivated in part by the \emph{strong cosmic censorship conjecture}. Roughly, the conjecture states that the maximal globally hyperbolic development of generic initial data for the Einstein equations is inextendible as a suitably regular Lorentzian manifold. 
Formulating a precise statement of the strong cosmic censorship conjecture is itself a challenge because one needs to make precise the phrases `generic initial data' and `suitably regular Lorentzian manifold'. Understanding the latter is where general relativity in low regularity and in particular (in-)extendibility results become significant.

 Christodoulou \cite{Christo1} established a $C^0$-inextendibility result for spacetimes satisfying the Einstein-scalar field equations within the class of spherically symmetric spacetimes. Likewise, the authors in \cite{GL} demonstrated the $C^0$-inextendibility of open FLRW spacetimes which are not Milne-like also within the class of spherically symmetric spacetimes. Moreover, they demonstrate, to the contrary, that Milne-like spacetimes  are always $C^0$-extendible (but might be $C^2$-inextendible).
Dafermos \cite{Dafermos03, Dafermos05} demonstrated the $C^0$-extendibility of the maximal globally hyperbolic development of solutions to the spherically symmetric Einstein-Maxwell-scalar field system arising from small perturbations of Reissner-Nordstr{\"o}m initial data. In fact, more recently, Dafermos and Luk \cite{DL} have given a proof, without symmetry assumptions, of the $C^0$ stability of the Kerr-Cauchy horizon. This gives firm evidence that the strong cosmic censorship censorship is false in the $C^0$ setting. The current suggestion for the statement of the strong cosmic censorship conjecture is to require inextendibility as a Lorentzian manifold with a continuous metric and Christoffel symbols locally in $L^2$ \cite{Christo2, DL}.

Still, there are some interesting new results establishing the $C^0$-inextendibility of spacetimes without any symmetry assumptions of the extension. The first such example was given by Sbierski \cite{Sbierski} with his proof of the $C^0$-inextendibility of the Schwarzschild spacetime. Further advancements have been made in this direction \cite{GLS}, and our second application adds to the list of these inextendibility results.

 A systematic study of general relativity in low regularity began with the influential paper \cite{CG} where the authors studied causal theory in spacetimes where the regularity of the metric was less than $C^2$. Since then further advancements have established that most of classical causal theory remains valid for $C^{1,1}$ spacetimes \cite{KSS, KSSV, M} and even the singularity theorems hold in this regularity class \cite{hawkingc11, penrosec11, hawkpenc11}. Therefore spacetimes with $C^{1,1}$ metrics can be seen as the threshold to where classical causality theory applies.

Once the regularity drops below $C^{1,1}$, causality theory departs significantly from classical theory. Example 1.11 in \cite{CG} shows that the push-up Lemma (i.e. $I^+(J^+(\Omega)) = I^+(\Omega)$) does not necessarily hold in $C^{0,\alpha}$ spacetimes, $\alpha \in (0,1)$. This led the authors to define the so-called bubbling regions for these spacetimes. From Example 1.11 it is readily seen that any causal curve from the origin to a point $p$ in the bubbling region must begin null on some interval. Thus any causal curve which maximizes the Lorentzian length between the origin and $p$ is null on an interval, hence not timelike. However, since these points have positive Lorentzian distance, it must be timelike on a set of non-zero measure, so it cannot be null. This deviates drastically from classical (at least $C^{1,1}$) theory where maximizers must be geodesics and hence either timelike or null.


On the other hand \cite[Corollary 1.17]{CG} shows that $C^{0,1}$ spacetimes (i.e. spacetimes with a Lipschitz continuous metric) do not admit bubbling regions. This leaves open the question of whether maximizers in $C^{0,1}$ spacetimes must be either timelike or null. In fact this question was posed in \cite{SS}. In this paper we prove the following theorem which answers the question affirmatively.

\begin{Theorem}\label{main}
	Let $(M,g)$ be a Lipschitz spacetime. If $p,q\in M $ with $q \in J^+(p)$, then any maximizing causal curve from $p$ to $q$ is either timelike or null. 
\end{Theorem}

As an application we show


\begin{Theorem}\label{inextend theorem}
Let $(M,g)$ be a smooth timelike geodesically complete spacetime. Then $(M,g)$ is $C^{0,1}$-inextendible.
\end{Theorem}

This provides a partial answer to a question raised in \cite{Sbierski} which asks whether timelike geodesically complete spacetimes are $C^0$-inextendible. Together with the corresponding result from \cite{GLS}, which states that smooth
timelike geodesically complete globally hyperbolic spacetimes are $C^0$-inextendible, this shows that if $(M_{\rm ext}, g_{\rm ext})$ is a $C^0$ extension of a timelike geodesically complete spacetime $(M,g)$, then $(M,g)$ is not globally hyperbolic and $g_{\rm ext}$ is not Lipschitz.

We also note that this establishes that strongly causal Lipschitz spacetimes are examples of \emph{regular} Lorentzian length spaces as introduced in \cite{KS}.

\section{Proofs of Theorems 1.1 and 1.2}
\label{sec:proof}

Our definitions for a $C^{k,\alpha}$ \emph{spacetime} and \emph{future causal curves} $\gamma \colon I \to M$ follow that of \cite{CG}. In particular we use Lipschitz causal curves which implies that the derivative $\dot{\gamma}$ of $\gamma$ exists almost everywhere and $\dot{\gamma}$ is locally in $L^\infty$. We say a Lipschitz curve is future causal if $\dot{\gamma}$ is future causal almost everywhere. A future causal curve is \emph{timelike} if $\dot{\gamma}$ is timelike almost everywhere and it is \emph{null} if $\dot{\gamma}$ is null almost everywhere. The \emph{Lorentzian length} of a future causal curve $\gamma \colon I\to M$ is the integral $L(\gamma) = \int_I \sqrt{-g(\dot{\gamma}, \dot{\gamma})}$ and $\gamma$ is said to be \emph{maximizing} or a \emph{maximizer} if $L(\gamma) \geq L(\lambda)$ for any future causal curve $\lambda$ whose endpoints agree with those of $\gamma$.

\proof[Proof of Theorem \ref{main}]
Suppose $\gamma \colon I \to M$ is a maximizing future directed causal curve from $p$ to $q$ which is not null. We will prove that $\gamma$ is timelike. Seeking a contradiction, suppose the set $N_I:=\{s\in I: \dot{\gamma}(s)\; \mathrm{exists\:and\:is\:null}\}\sse I$ has positive measure. Below we will construct another causal curve from $p$ to $q$ which is longer than $\gamma$, and hence contradicting the fact that $\gamma$ is a maximizer.

To do this, we first want to localize the situation. By compactness we may cover $I$ by finitely many open intervals $I_k$ (half open intervals on the endpoints of $I$) such that each $\gamma |_{I_k}$ is contained in a relatively compact chart domain $(U_k, \varphi_k)$ on  which $g(\partial_0^{\varphi_k}, \partial_0^{\varphi_k})<c_k<0 $. We are now going to show that for at least one of these $I_k $, we have $0< \mu (N_{I_k}) < \mu (I_k)$ where $\mu$ is the usual Lebesgue measure on $\R$. That is we will show there is a $k$ such that  $\gamma |_{I_k}$ is causal but neither  timelike nor null. Assume for the moment that $\mu(N_{I_j})=0$ for some $j$. Then, since the intersection of the neighbouring intervals $I_{j-1}$ and $I_{j+1}$ with $I_j$ must be non-empty and open, either one of those has the desired property or $\mu(N_{I_{j-1}})=\mu(N_{I_{j-1}})=0$. The existence of a suitable $I_k$ now follows by induction and noting that $\mu(N_I)\neq 0$. If instead $\mu(N_{I_j})=\mu (I_j)$, one proceeds the same way, using $\mu(N_I)\neq \mu(I)$ in the end. 

This shows that we may assume w.l.o.g. that $\gamma (I)$ is contained in such a chart domain $(U,\varphi)$. By reparametrizing $\gamma $ we may further assume that $\dot{\gamma}(0)$ exists and is timelike. Using a linear change of coordinates corresponding to a Gram-Schmidt orthogonalization process of $\{ \partial_0^\varphi,\dots ,\partial_{n-1}^\varphi\}|_{\gamma(0)}$ and a translation we get new coordinates $\psi $ on $U$ for which $\partial_0^\psi \propto \partial_0^\varphi$ (and hence $g(\partial_0^\psi, \partial_0^\psi)<c<0 $) on $U$, $\gamma^\psi(0)=0$ and $g^\psi(0)=\eta $ where $\eta$ is the Minkowski metric.

To sum up, we need only consider the case $M=\R^n$, $\gamma \colon[a,b] \to U\sse \R^n$, $\gamma (0)=0$, $\gamma $ is differentiable at $0$ and $\dot{\gamma}(0)$ is timelike, $\mu (N_{[a,0]}) >0$ (if instead $\mu (N_{[0,b]}) >0$ one just needs to reverse the time orientation), $g(0)=\eta $ and $\partial_0 $ (uniformly) timelike on $U$.

We first look at $\gamma_1 :=\gamma|_{[a,0]}$. As in \cite[Lem.~1.15]{CG}, given a Lipschitz function $f\colon [a,0]\to [0,\infty)$, we define a new Lipschitz curve $\Gamma_1\colon [a,0] \to \R^n$ by $\Gamma_1^\mu(s) := \gamma_1^\mu(s) + \epsilon f(s)T^\mu$  where $T^\mu$ is defined via $\partial_0 = T^\mu\partial_\mu$ (i.e. $T^0 = 1$ and $T^\mu = 0$ for $\mu \neq 0$). Then the proof of \cite[Lem.~1.15]{CG} shows that for a Lipschitz metric one can find a specific $f\in C^{0,1}([a,0])$ such that $\Gamma_1 \sse U$, $\Gamma_1(a)=\gamma_1(a)$, $\Gamma_1(0)=\big(\epsilon f(0), 0, \dotsc, 0\big)$
and 
\begin{equation} \label{eq:CG bound}
g_{\Gamma_1}(\dot{\Gamma}_1,\dot{\Gamma}_1)\leq g_{\gamma_1}(\dot{\gamma}_1,\dot{\gamma}_1)-\frac{\eps	}{2}
\end{equation}
a.e. on $[a,0]$ for all $\eps $ less than some $\eps_0$. This gives the following estimate for the lengths of $\gamma_1$ and $\Gamma_1$:
\begin{align}
\label{eq: length estimate 1}
L(\Gamma_1)&=\int_a^0 \sqrt{	-g_{\Gamma_1}(\dot{\Gamma}_1,\dot{\Gamma}_1)} = \int_{N_{[a,0]}} \sqrt{	-g_{\Gamma_1}(\dot{\Gamma}_1,\dot{\Gamma}_1)} +\int_{[a,0]\setminus N_{[a,0]}} \sqrt{-g_{\Gamma_1}(\dot{\Gamma}_1,\dot{\Gamma}_1)}
\\
&= \frac{\mu (N_{[a,0]})}{\sqrt{2}} \sqrt{\eps}+ \int_{[a,0]\setminus N_{[a,0]}} \sqrt{-g_{\Gamma_1}(\dot{\Gamma}_1,\dot{\Gamma}_1)}
\\
&\geq \frac{\mu (N_{[a,0]})}{\sqrt{2}} \sqrt{\eps}+L(\gamma_1).
\end{align}
Note that this not only shows that $L(\Gamma_1)\geq L(\gamma_1)$ but more importantly that the length difference is bounded from below by an expression that scales like $\sqrt{\eps}$ as $\eps \to 0$.

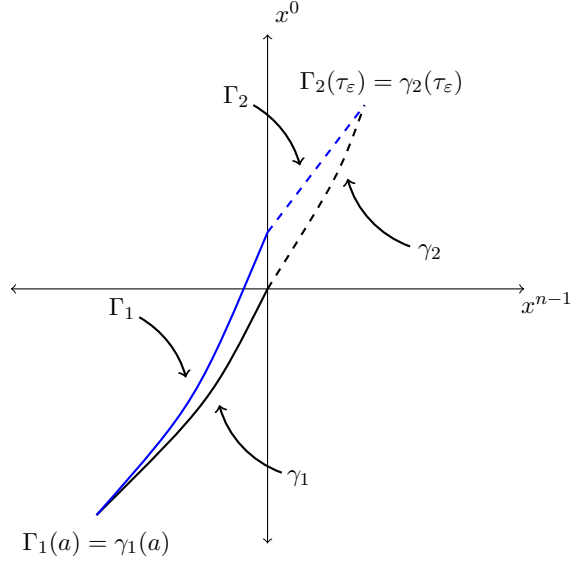
\begin{figure}[ht] 
\[
\begin{tikzpicture}[scale = .75]

\draw [<->] (0,-6.5) -- (0,2.5);
\draw [<->] (-4.5,-2) -- (4.5,-2);

\draw (.3550,2.90) node [scale = .85] {$x^0$};

\draw (4.90, -2.25) node [scale = .85] {$x^{n-1}$};

\draw [->] [thick] (-2.25,-2.5) arc [start angle=60, end angle=15, radius=50pt];

\draw (-2.55,-2.35) node [scale = .85] {$\Gamma_1$};

\draw [->] [thick] (-.25,1.25) arc [start angle=60, end angle=15, radius=50pt];

\draw (-.55,1.40) node [scale = .85] {$\Gamma_2$};

\draw [thick] (-3, -6) .. controls (-1,-4) .. (0, -2);

\draw [thick,blue] (-3,-6) .. controls (-1.25, -4) .. (0, -1);

\draw (-3,-6.5) node [scale = .85] {$\Gamma_1(a) = \gamma_1(a)$};

\draw [thick, dashed] (0,-2) .. controls (1.25 ,0 ) .. (1.70, 1.25);

\draw [thick,blue, dashed] (0,-1) -- (1.70,1.25);	

\draw (2,1.65) node [scale = .85] {$\Gamma_2(\tau_\eps) = \gamma_2(\tau_\eps)$};	

\draw [->] [thick] (.25,-5.25) arc [start angle=250, end angle=195, radius=50pt];

\draw (.55,-5.35) node [scale = .85] {$\gamma_1$};	

\draw [->] [thick] (2.5,-1.25) arc [start angle=250, end angle=195, radius=50pt];

\draw (2.85,-1.35) node [scale = .85] {$\gamma_2$};

\end{tikzpicture}
\]
\caption{\small{The causal curve formed by concatenating $\Gamma_1$ and $\Gamma_2$ has Lorentzian length greater than that of $\gamma|_{[a, \tau_\eps]}$. Hence $\gamma$ cannot be maximizing.}}
\label{picture}
\label{figure}
\end{figure}

We now turn to $\gamma|_{[0,b]}$ and assume that $\gamma|_{[0,b]}$ is parametrized by the $x^0$-coordinate. The chain rule ensures that $\dot{\gamma}(0)$ exists under this reparameterization and hence remains timelike.
 Since $\Gamma_1(0)\neq \gamma (0)$ we have to find $\tau_\eps>0$ 
and a future directed causal curve $\Gamma_2$ from $\Gamma_1(0)=(\eps f(0),0)$ to $\gamma(\tau_\eps)=\big(\tau_\eps, \bar{\gamma}(\tau_\eps)\big)\in \R \times \R^{n-1}$ (see figure \ref{picture}) and such that for $\eps $ small enough $L(\Gamma_1)+L(\Gamma_2)>L(\gamma_1)+L(\gamma_2)$ where $\gamma_2 := \gamma|_{[0,\tau_\eps]}.$ It should be noted that the segment $\gamma_2$ of $\gamma$ depends on $\tau_\eps$ and hence depends on $\eps$ itself. It suffices to show $L(\Gamma_1)>L(\gamma_1)+L(\gamma_2)$. And using \eqref{eq: length estimate 1} we see that this holds if $L(\gamma_2)$ has an upper bound that scales like $\eps^k$ for some $k>\frac{1}{2}$ as $\eps \to 0$. Below we will show this is true for $k = 1$.

Since $\gamma $ is differentiable at $s=0$, Taylor's theorem gives
\begin{equation}\label{eq:taylor for gamma}
\gamma(s)=\gamma(0)+\dot{\gamma}(0)s+h(s)s=(\dot{\gamma}(0)+h(s))s,
\end{equation}
where $h(s)\to 0$ as $s\to 0$. Since we further assumed $\dot{\gamma}(0)=:(v^0,\bar{v})$ to be timelike and $g(0)=\eta$, we can choose $\alpha > 1$ such that $\frac{v^0}{|\bar{v}|_e}>\alpha $ where $|\cdot|_e$ is the usual Euclidean norm. Together with \eqref{eq:taylor for gamma} this shows that there exists $s_0>0$ such that $s=\gamma^0(s)>\alpha |\bar{\gamma}(s)|_e$ for all $s<s_0$. Choose $1 < \beta < \alpha$ and let $C_{\beta, d}$ denote the future cone with slope $\beta$ that has its tip
in $(d,0)$, i.e., $C_{\beta, d}=\{(\beta |\bar{x}|_e+d,\bar{x}): \bar{x}\in \R^{n-1}\}$. Since $\beta >1$ and $g(0)=\eta$ the continuity of $g$ allows us to find a small neighbourhood $V$ of $0$ such that $C_{\beta, d}\cap V\setminus \{(d,0)\} \sse I^+((d,0))$ for all small $d > 0$. We have $(s,\bar{\gamma}(s))\in C_{\beta,s-\beta |\bar{\gamma}(s)|_e}\cap V \sse I^+((s-\beta |\bar{\gamma}(s)|_e,0))$ for small $s$. Since $s>\alpha |\bar{\gamma}(s)|_e$ we have $s-\beta |\bar{\gamma}(s)|_e>s(1-\frac{\beta}{\alpha})$ which shows that $(s,\bar{\gamma}(s))\in I^+((s(1-\frac{\beta}{\alpha}),0))$ for small $s$. Thus $\gamma(\tau_\eps)=(\tau_\eps,\bar{\gamma}(\tau_\eps))\in I^+((\eps f(0),0))$ for $\tau_\eps=\frac{f(0)}{1-\frac{\beta}{\alpha}}\eps$ if $\eps $ is small enough. Therefore we have demonstrated that we can construct $\Gamma_2$ from $\Gamma_1(0)$ to $\gamma(\tau_\eps)$ for small $\eps$. Finally, for $L(\gamma_2)$ an estimate from the proof of Thm.~3.3 in \cite{GLS} shows that, if $\gamma_2 \sse V$, where $V$ is a neighborhood of $0$ on which $|g_{\mu \nu}-\eta_{\mu \nu}|<\delta $, then
\begin{equation}
L(\gamma_2)\leq \tau_\eps \sqrt{1+\delta +4 (n-1)^2 \delta}=\eps \, \frac{f(0)}{1-\frac{\beta}{\alpha}} \sqrt{1+\delta +4 (n-1)^2 \delta}.
\end{equation}
So $L(\gamma_2)$ is indeed bounded from above by a term of order $\eps $. Now choose $\eps$ small enough so that 
\[
\frac{\mu (N_{[a,0]})}{\sqrt{2}} \sqrt{\eps} \geq \eps \, \frac{f(0)}{1-\frac{\beta}{\alpha}} \sqrt{1+\delta +4 (n-1)^2 \delta},
\]
then we have
\begin{align*}
L(\Gamma_1) + L(\Gamma_2) &\geq L(\Gamma_1) \geq L(\gamma_1) + \frac{\mu (N_{[a,0]})}{\sqrt{2}} \sqrt{\eps}\\
  &\geq L(\gamma_1) + \eps \, \frac{f(0)}{1-\frac{\beta}{\alpha}} \sqrt{1+\delta +4 (n-1)^2 \delta}
\\
&\geq L(\gamma_1) + L(\gamma_2),
\end{align*}  
and so we are done. \qedhere

\medskip

\noindent\emph{Remark.} Our proof also shows if $\gamma$ is a maximizing null curve between two points, then $\gamma$ does not contain a single timelike tangent. This result is also obtained in \cite[Theorem 18]{M2} under much weaker differentiability assumptions.


\proof[Proof of Theorem \ref{inextend theorem}]
Seeking a contradiction, suppose such a Lipschitz extension $(M_{\rm ext}, g_{\rm ext})$ exists. Following the proof of \cite[Theorem 3.3]{GLS}, the future causal curve $\alpha$ is a maximizer from $q \in M$ to $p \in \partial^+M$ within a small globally hyperbolic set $V$ containing $q$ and $p$. Since $q$ and $p$ are timelike separated within $V$, the curve $\alpha$ is timelike by Theorem \ref{main}. Since $\alpha$ is a timelike maximizer and leaves $M$, the portion of $\alpha$ that lies in $M$ is an inextendible (within $M$) timelike geodesic. Then timelike geodesic completeness of $(M,g)$ implies $L(\alpha) = \infty$, contradicting $L(\alpha)=d_V(q,p)<\infty$. 
\qed

\medskip

Lastly we remark that in \cite{SS}, it was suggested that if maximizers in $C^1$ spacetimes are timelike, then it may be possible to apply the du Bois-Reymond-trick to show that the maximizers must be $C^2$ timelike geodesics. However it seems like the du Bois-Reymond-trick does not apply for timelike maximizers in the Lorentzian setting. This is because one can not ensure that the variation of the maximizer is timelike, since the tangent of a Lipschitz timelike curve may still come arbitrarily close to being null. This is not an issue in the Riemannian setting since there is no causal distinction between the curves.

\medskip

\noindent{\em Acknowledgements.} 
Melanie Graf is the recepient of a DOC-fellowship of the Austrian Academy of Sciences. This work was completed as part of her research stay at the University of Miami made possible by a scholarship of the Austrian Marshall Plan Foundation. This work was also partially supported by project P28770 of the Austrian Science Fund FWF. The authors are grateful to Greg Galloway for helpful comments. The authors would also like to thank Clemens S\"{a}mann and Michael Kunzinger for suggesting the problem and Ettore Minguzzi for bringing \cite{M2} to their attention. We also thank an anonymous referee for pointing out the problems with the du Bois-Reymond-trick.

\end{document}